\documentclass[acmlarge,screen]{acmart}
\usepackage{tabularx}

\AtBeginDocument{%
  \providecommand\BibTeX{{%
    \normalfont B\kern-0.5em{\scshape i\kern-0.25em b}\kern-0.8em\TeX}}}

\copyrightyear{2023} 
\acmYear{2023} 
\setcopyright{acmlicensed}\acmConference[CHI '23]{Proceedings of the 2023
CHI Conference on Human Factors in Computing Systems}{April 23--28,
2023}{Hamburg, Germany}
\acmBooktitle{Proceedings of the 2023 CHI Conference on Human Factors in
Computing Systems (CHI '23), April 23--28, 2023, Hamburg, Germany}
\acmPrice{15.00}
\acmDOI{10.1145/3544548.3581419}
\acmISBN{978-1-4503-9421-5/23/04}

\usepackage{xcolor}
\usepackage{color}
\usepackage{multirow}
\usepackage{makecell}
\usepackage{subfig}

\begin{document}

\title[MemoAnalyzer]{``Ghost of the past'': Identifying and Resolving Privacy Leakage of LLM's Memory Through Proactive User Interaction}


\author{Shuning Zhang}
\authornotemark[1]
\email{Zhang.sn314@gmail.com}
\affiliation{
    \institution{Institute for Network Sciences and Cyberspace, Tsinghua University}
    \city{Beijing}
    \country{China}
}
\author{Lvmanshan Ye}
\authornote{Both authors contributed equally to this work.}
\affiliation{
    \institution{Shanghai Jiao Tong University}
    \city{Shanghai}
    \country{China}
}
\author{Xin Yi}
\email{yixin@tsinghua.edu.cn}
\authornote{This is the corresponding author.}
\affiliation{
    \institution{Institute for Network Sciences and Cyberspace, Tsinghua University}
    \city{Beijing}
    \country{China}
}
\affiliation{
    \institution{Zhongguancun Laboratory}
    \city{Beijing}
    \country{China}
}

\author{Jingyu Tang}
\affiliation{
    \institution{Huazhong University of Science and Technology}
    \city{Wuhan}
    \country{China}
}

\author{Bo Shui}
\affiliation{
    \institution{Shenzhen International Graduate School, Tsinghua University}
    \city{Shenzhen}
    \country{China}
}

\author{Haobin Xing}
\affiliation{
    \institution{Tsinghua University}
    \city{Beijing}
    \country{China}
}

\author{Pengfei Liu}
\email{pengfei@sjtu.edu.cn}
\affiliation{
    \institution{School of Electronic, Qingyuan, Shanghai Jiao Tong University}
    \city{Shanghai}
    \country{China}
}

\author{Hewu Li}
\affiliation{
    \institution{Tsinghua University}
    \city{Beijing}
    \country{China}
}

\begin{abstract}

Memories, encompassing past inputs in context window and retrieval-augmented generation(RAG), frequently surface during human-LLM interactions, yet users are often unaware of their presence and the associated privacy risks. To address this, we propose MemoAnalyzer, a system for identifying, visualizing, and managing private information within memories. A semi-structured interview(N=40) revealed that low privacy awareness was the primary challenge, while proactive privacy control emerged as the most common user need. MemoAnalyzer uses a prompt-based method to infer and identify sensitive information from aggregated past inputs, allowing users to easily modify sensitive content. Background color temperature and transparency are mapped to inference confidence and sensitivity, streamlining privacy adjustments. A 5-day evaluation(N=36) comparing MemoAnalyzer with the default GPT setting and a manual modification baseline showed MemoAnalyzer significantly improved privacy awareness and protection without compromising interaction speed. Our study contributes to the growing field of privacy-conscious LLM design, offering insights into user-centric privacy protection for Human-AI interactions.

\end{abstract}

\keywords{Large Language Models, Memory, Privacy Inference, Privacy Awareness}


\maketitle

\section{Introduction}

The widespread use of Conversational Agents (CAs) based on Large Language Models (LLMs) has facilitated natural language communication while simultaneously posing significant challenges to users' privacy~\cite{gao2024taxonomy}. To enhance user experience and minimize repeated input of task-related background information, LLM service providers, such as OpenAI, have implemented various memorization techniques. These methods involve deducing and storing user information based on natural language interactions\cite{zhong2024memorybank,wang2024augmenting}. The powerful inference capabilities of LLMs, combined with increasing user data disclosure~\cite{zhang2024s,staabbeyond}, embed significant private information in memory traces that persist indefinitely unless explicitly deleted by the userthemselves~\footnote{\url{https://openai.com/index/memory-and-new-controls-for-chatgpt/}, accessed by Sep 12th, 2024}. As interactions with LLM-powered chatbots become more frequent across all aspects of life, the threat of privacy-invasive chatbots has risen, particularly concerning memory.  Moreover, memory generation and usage in LLMs often lack transparency~\cite{zhong2024memorybank}, frequently occurring without user consent or awareness. 


\textbf{Long-term memory operates using a retrieval-augmented generation (RAG-based) method, while short-term memory retains past user input in the context window.} These approaches mimic human memory mechanisms~\cite{repovvs2006multi} and together form the LLM’s “memory” of user input. Both memorized content could be leveraged for training, posing significant privacy leakage risks~\cite{staab2023beyond}. This opacity leads to inadvertent user consecutive contributions to memory systems, heightening privacy risks~\cite{pan2020privacy} and vulnerability to membership inference attacks~\cite{staab2023beyond}. For example, You have told ChatGPT about your preferences for work hours and life balance. Recently, you have asked ChatGPT for help about the difficulties you encounter at work. When brainstorming the possibility of changing jobs or starting a business. ChatGPT may infer dissatisfaction with your job and plans for a career change based on prior conversations.


This research builds on previous classifications \cite{zhang2024survey} to analyze memory-related risks and corresponding countermeasures across memory generation and usage phases. Our focus is on privacy leakage risks \cite{kim2024propile} and user inference attacks \cite{kandpal2023user}, both critical threats in LLMs \cite{lukas2023analyzing, kandpal2023user}. Two main privacy risks arise from user inputs when transferred to memory: (1) individual inputs or long-term memories may contain sensitive information, and (2) the aggregation of these past inputs and memories may lead to the exposure of sensitive data. When utilized for model training or fine-tuning, these inputs pose a significant privacy threat by potentially enabling the model to infer and expose personal information~\cite{chen2024combating,brown2022does}.

The past input and memory consisting sensitive information, as well as the inferred private information both presented risks of privacy leakage~\cite{kim2024propile,lukas2023analyzing}. The persistent use of such memory intensifies these risks, yet privacy implications remain under-explored, with limited research on risk categorization and mitigation. Hence, we aim to answer this research question: \textbf{How to design a transparent and controllable notification technique which timely triggers participants' awareness and mitigates potential risks in the memories of LLMs-based CA?}

To explore these questions, we first conducted a semi-structured formative interview study (N=40) to examine users' privacy perceptions towards LLM memory systems and their expectations for efficient memory management. The results showed that most users were unaware of the existence of memory systems, particularly long-term RAG-based memory, with only 5 out of 40 participants demonstrating an understanding of long-term mechanisms. Even users with frequent usage held misconceptions, such as believing memory was limited to a specific dialogue or could be shared with other users. After our explanation, participants expressed the need for transparent, controllable designs for visualizing and modifying the memory mechanism. We found that current LLM products already retain some personal privacy information in their memory, which accumulates over time. However, the inference process for this private information remains opaque, and users often only realize privacy risks after a delay, when tracing the original input becomes significantly more difficult.

To address the challenges identified in the formative study, we developed MemoAnalyzer, a pop-up browser plug-in that visualizes inferred private information and enables users to modify it easily. MemoAnalyzer appears after each interaction, distinguishing private information based on inference confidence and sensitivity~\cite{10561732, liu2023chatbots}. Inference confidence is represented through varying opacity levels, while sensitivity is indicated by color, with red signaling highly sensitive information and blue indicating lower sensitivity. When users click on a piece of private information, MemoAnalyzer displays the relevant past inputs and memories, highlighting the keywords that contributed to the inference. This helps users quickly identify and modify the terms that contributed to the inference. By utilizing prompt-based inferences, MemoAnalyzer provides users with proactive control over their inputs and memory data. Since it only performs inference without training on the data until users take action, privacy risks are minimized at this stage. Once users modify or delete their data, any potential risk from future model training is effectively mitigated.

A five-day in-lab evaluation study (N=36) demonstrated the effectiveness of MemoAnalyzer compared to \textit{GPT} implementation and \textit{Manual} management in three typical types of tasks: work-related, life-related and academic-related \cite{zhang2024s}. MemoAnalyzer was preferred for its comparable time efficiency, superior privacy protection, and enhanced user experience regarding perceived control, transparency, trust, and overall preference. Users particularly appreciated its flexibility, control-ability, and transparency. We envision MemoAnalyzer to be a superior solution for managing personal memory in LLMs, especially concerning sensitive information.

In summary, this work made three key contributions:
\begin{itemize}
    \item We unveiled users generally lacked the timely and clearly awareness of the long-term memory mechanism in contrast to the short-term context memory through a interview study (N=40). 
    \item We proposed MemoAnalyzer, a technique notifying users' privacy risk, enabling users to selectively control their private information. MemoAnalyzer facilitated the collaborative privacy information management where users indicate their preference.
    \item We evaluated MemoAnalyzer in an user study (N=36) compared with GPT and Manual settings, where MemoAnalyzer were favored for its comparable speed, higher privacy protection capability and higher user experience. It also enabled users to control their privacy more transparently.  
\end{itemize}

 
\section{Related Works}

We first introduced the privacy risk in LLM-based CAs. Then we detailed the memory mechanisms of LLMs and the potential privacy risk. Finally, we categorized the privacy awareness of end-users in human-AI interaction. 

\subsection{Privacy Risk in LLM-based CAs}\label{sec:rw_privacy_risk_llm}

With a service targeting conversational assistants, we detail privacy challenges in LLM-based CAs memorization. To optimize conversational performance, LLMs inherently require vast amounts of data for their training, often encompassing user interaction data \cite{pahune2023several}. However, a side effect of LLMs is the unintentional memorization of the training data, which also contain user input data, including personally identifiable information (PII) \cite{brown2022does,peris2023privacy}, which might also be included in the generated output. For example, ChatGPT, even with safety precautions, can inadvertently disclose PII through specifically crafted prompts.

Users engage with LLM-based CAs through natural language, which is traditionally reserved for human-to-human communication. This can lead them to perceive these agents as human-like. Studies suggested that anthropomorphizing can increase user information disclosure \cite{ischen2020privacy, kim2012anthropomorphism}. Anthropomorphizing can inflate users' perceptions of the CAs' competencies, fostering undue confidence, trust, or expectations in these agents \cite{kim2012anthropomorphism,zlotowski2015anthropomorphism}. With more trust, users might be more inclined to share private information, even in contexts typically associated with sensitive personal information \cite{kim2012anthropomorphism,waldman2018privacy,zlotowski2015anthropomorphism}. Anthropomorphization may amplify the risks of users yielding effective control by trusting CAs unquestioningly. Moreover, more private information may be revealed when CAs leverage psychological effects, such as nudging or framing \cite{weidinger2021ethical}. 

Generalized to AI technologies, the past literature \cite{lee2024deepfakes} classified the risks as invasion risks \cite{solove2005taxonomy,otrel2010taking,ping2018automatic,milmo2021amazon}, data collection risks \cite{solove2005taxonomy}, data dissemination risks \cite{ayyub2018india,levin2017new,burgess2021biggest} and data processing risks \cite{solove2005taxonomy,simonite2018facebook,pannett2022german,pearl2010faces,stark2021physiognomic}. Invasion risks encompass a range of activities that intrude upon an individual's personal space or solitude. Intrusion risks encompass actions that disturb one's solitude in physical space \cite{solove2005taxonomy}, which include personalized ads. Besides, surveillance is common with the support of AI technologies \cite{otrel2010taking,ping2018automatic}, and the ubiquity of sensors \cite{milmo2021amazon}. Data collection risks ``create disruption based on the data gathering process'' \cite{solove2005taxonomy}, which exacerbated surveillance risks \cite{solove2005taxonomy}, further exacerbating surveillance risks. Data processing risks result from the use, storage and manipulation of personal data \cite{solove2005taxonomy,simonite2018facebook}. 



\subsection{Memory in LLMs}\label{sec:rw_interaction_privacy_risk}

With the increasing complexity of human-AI interaction~\cite{yang2020re} and the tasks~\cite{amershi2019guidelines}, memory becomes significant during the interaction~\cite{huang2023memory,bae2022keep}. Current LLMs remain opaque about the usage of memory~\cite{huang2023memory} and the context~\cite{wu2023brief,zhao2022improving}. To address these challenges, researchers have explored various strategies to manage memory more effectively and transparently. Some proposed techniques for retaining the persona of chatbots~\cite{li2016persona,zhang2018personalizing}. Other methods guarantee the responses generated are contextually appropriate, such as summarizing~\cite{wang2023recursively} and refinement~\cite{zhong2022less}, aiming to minimize redundancy while maintaining essential information. Relevant memories can be retrieved utilizing information retrieval techniques to contextualize current inputs to AI~\cite{xu2022,bae2022keep}. One of the critical aspects of privacy risks in LLM-based systems is the memory mechanism these models employ to retain and utilize user data across interactions. Memory in LLMs can be broadly categorized into short-term memory, which involves retaining user input for the duration of a session, and long-term memory, which stores user interactions across multiple sessions to enhance continuity and personalization ~\cite{Zhong_Guo_Gao_Ye_Wang_2024}. Recent studies have highlighted the opacity and complexity of memory mechanisms in LLMs, which often operate without explicit user consent or understanding ~\cite{10234764}. For example, research showed that users are typically unaware of how LLMs store and use their data, leading to significant privacy concerns, especially regarding the potential for long-term retention of sensitive information ~\cite{abs-2311-08719}. 

This lack of transparency raises ethical concerns and poses substantial risks of data breaches and unauthorized access to private information ~\cite{staab2024beyond}. Researchers have proposed various strategies for managing memory in LLMs to address these challenges, focusing on enhancing transparency and user control. One approach involves using external memory management systems that allow users to access, modify, or delete stored data proactively ~\cite{luo-etal-2022-readability}. These systems often employ visualization techniques to help users understand what data has been retained and how it might be used in future interactions ~\cite{10.1145/3586182.3615796}. For instance, Huang et al. developed a memory sandbox tool that provides a transparent interface for managing conversational histories, enabling users to selectively edit or remove entries that contain sensitive information ~\cite{NEURIPS2023_420678bb}. Moreover, advancements in prompt-based memory management have been explored to facilitate better control over the data retained by LLMs. This involves using structured prompts to guide the model in identifying and managing relevant information while minimizing the retention of unnecessary or sensitive data ~\cite{zhong-etal-2022-less}. Recent implementations of these techniques have demonstrated their effectiveness in reducing privacy risks and enhancing user trust by providing more granular control over memory retention and usage ~\cite{liang-etal-2022-modular}. 

Despite these efforts, the process of ``remembering'' remains complex or machines~\cite{liu2023think,wang2023recursively}, which is even harder for humans to take control. Users often lack a clear understanding of how generative AI and conversational agents handle memories~\cite{huang2023memory}. Current tools primarily focus on accessing and editing chat histories to manage conversational memories~\cite{huang2023memory,openai2024memory}. 


\subsection{User-centered Privacy in Human-LLM Interaction}

Users' perception of privacy critically influences their disclosure behaviors and the potential for privacy leakage in human-LLM interactions. However, due to LLMs' opaque privacy management mechanisms, users often lack sufficient privacy awareness, undermining their ability to make informed decisions about personal data disclosure.

To address this pervasive issue, researchers have proposed various countermeasures within the "notice and control" paradigm. Yet, these measures frequently encounter significant limitations. Non-salient privacy notices fail to effectively capture users' attention, especially when concealed behind hyperlinks or embedded in click-wrapped agreements. For instance, Cate~\cite{cate2016failure} highlighted that on Yahoo's website in 2002, a mere 0.3\% of users read the click-wrapped privacy policy, a figure that increased to only 1\% after a public privacy controversy~\cite{hansell2002compressed}. Similarly, in an experiment with a fictitious search engine, none of the 120 participants accessed the privacy policy link~\cite{groom2011reversing}. In another study, only 26\% of users joining a simulated social network viewed the policies~\cite{groom2011reversing}, and in a survey scenario, just 20.3\% of participants clicked to view privacy information~\cite{steinfeld2016agree}.

Enhancing user control and transparency is therefore essential for fostering trust and safeguarding privacy in AI systems, including LLMs. Privacy-preserving techniques that empower users to modify or delete their personal information can substantially mitigate privacy risks~\cite{doi:10.1126/science.aaa1465}. Transparent AI systems that clearly communicate how data is collected, processed, and stored further strengthen user trust and promote responsible AI practices~\cite{doshivelez2017towards}. Adopting a user-centric approach to privacy design—emphasizing effective notice and granular control over personal data—has gained significant attention. Tools such as interactive privacy dashboards and real-time data usage notifications enable users to make informed decisions about their privacy~\cite{10.1145/3172944.3172961}. Research indicates that users are more inclined to trust and engage with AI systems that provide clear explanations of data handling and allow for easy adjustments to privacy settings~\cite{doi:10.1080/10447318.2020.1741118}. However, these systems lacked the analysis into the memory mechanism. Thus, this paper initiated the first study to visualize the privacy and enable users to modify the private information in memory.

\section{Study 1: Examining Users' Awareness and Practice using LLM's Memory}\label{sec:study_1}

In this section, we conducted an interview study to explore users' practices and privacy concerns regarding memory during their interactions with LLMs.




\subsection{Study Design}

The experiment was conducted through semi-structured interviews. The session began with questions about participants' demographics. Participants were then asked about their knowledge and experiences with the memory systems in LLM products. We then divided the memory usage into three stages according to the past literature~\cite{hu2024hiagent} and OpenAI's official introduction\footnote{https://openai.com/index/memory-and-new-controls-for-chatgpt/}: memory generation, memory management and memory usage. For each stage, we explained the relevant implementation and usage mechanisms, then asked participants to discuss the advantages and dis-advantages of these mechanisms. After gathering their subjective opinions, we inquired about their expectations for each memory stage. The full interview script is provided in Section~\ref{app:question_study1} of the appendix. Experimenters also posed additional open-ended questions if any unexpected insights emerged during the interviews.

\subsection{Recruitment and Participants}

\begin{figure}[!htbp]
    \includegraphics[width=0.4\textwidth]{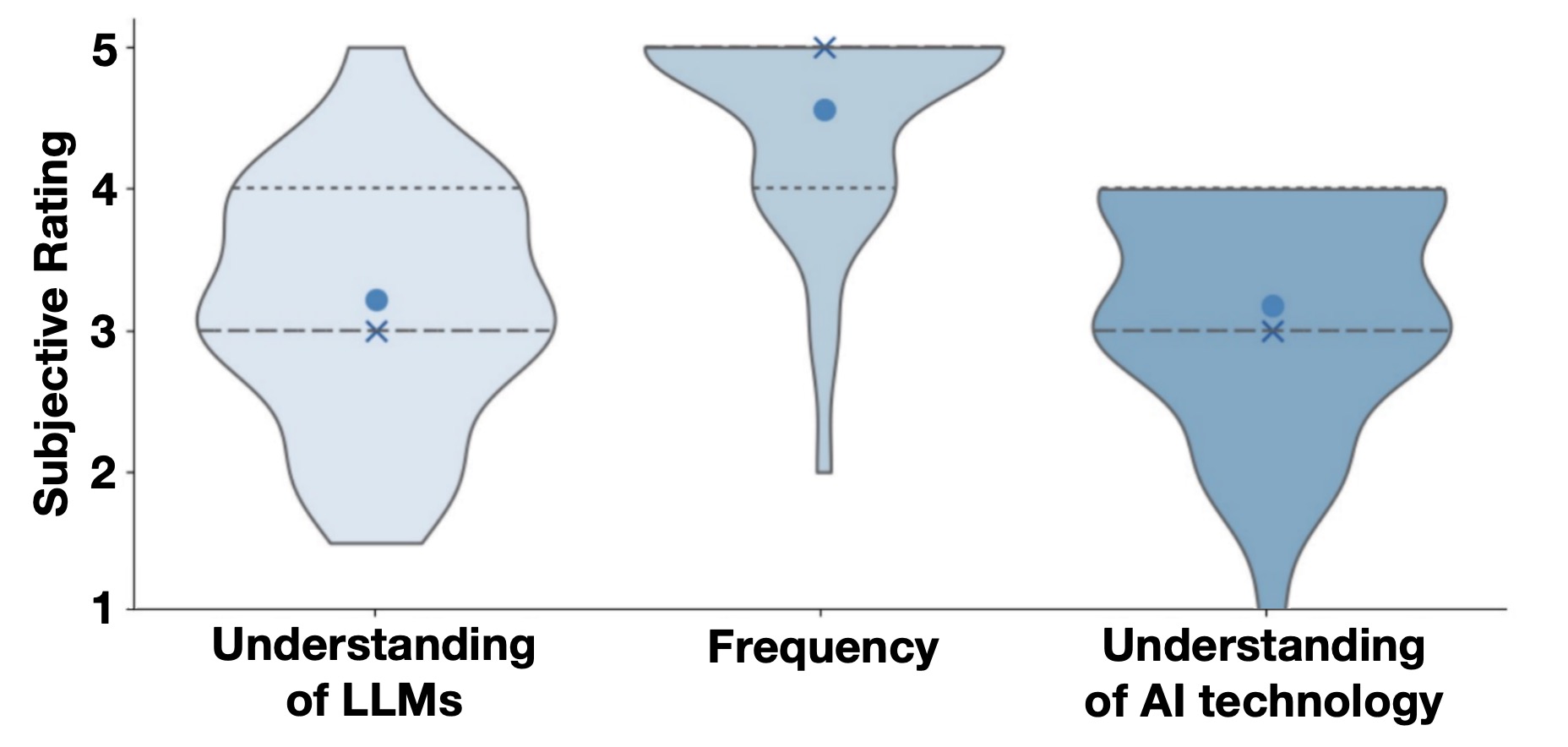}
    \caption{Participants' familiarity and usage frequency towards AI (5: most familiar and frequent, 1: least familiar and frequent). The cross sign indicated the median and the square sign indicated the mean.}
    \label{fig:study1_frequency}
\end{figure}

This IRB-approved study recruited 40 Chinese participants (13 males, 27 females) with a mean age of 22.6 (SD=2.1) in XX campus (anonymized for submission) through snowball sampling \cite{goodman1961snowball}. To ensure the diversity of participants, we distributed the questionnaire in different chat groups in different hours of a day. We continued the recruiting while limiting the participants each day until the results saturated following saturation theory \cite{fusch2015we}. After saturation, we continued to recruit another 3 participants. Five participants were with high school education degree, 16 were with bachelor education degree, 17 were with master degree and 2 were with Ph.D. degree. Participants self-rated their familiarity towards the LLMs and AI, as well as their usage frequency, the result of which was shown in Figure~\ref{fig:study1_frequency}. 7 participants were from design background and 6 were from computer science background, with no one from security and privacy background and other from other backgrounds. 39 participants have used ChatGPT \footnote{\url{https://chatgpt.com/}}, 19 participant have used Kimi Chat \footnote{\url{https://kimi.moonshot.cn/}} and 12 participants have used WenXinYiYan \footnote{\url{https://yiyan.baidu.com/}}. All participants reported having used LLMs at least once. Each participant who completed the experiment received 90 RMB as compensation.

\subsection{Procedure}

The experiment was conducted via Online Meeting\footnote{\url{https://meeting.tencent.com/}}. Participants were first briefed on the study and asked to provide informed consent. They were free to withdraw at any point during the experiment process. The entire experiment process lasted about 50 minutes on average and all recorded sessions were transcribed for analysis.

\subsection{Analysis Methods}
We used thematic analysis \cite{braun2012thematic} to qualitatively code the results. As the study followed grounded theory, the codebook was iteratively refined throughout the process. The first two authors jointly coded all data, engaging in periodic discussions to resolve disagreements. The coding process incorporated open coding \cite{khandkar2009open}, axial coding \cite{kendall1999axial} and inductive coding \cite{chandra2019inductive}. Initially, open coding was used to identify the primary set of codes, which were then grouped into axial codes and broader themes. Because of the inductive and iterative nature of the experiment, guided by the past literature \cite{mcdonald2019reliability}, agreement score is not suitable for this scenario. We intended not to report the agreement score. We calculated the frequency of utterance after qualitative coding.

\subsection{Results}\label{sec:study1_result}
We first revealed the cognitive gap of participants, then presented users' expectations towards future improvement. We detailed three findings shortened as F1 to F3.

\subsubsection{F1: Memory Mechanism is Opaque}


The process of memory generation within the system is problematic due to its opacity. Without explicit notification, 30/40 of users were un-aware of the notification when memory and context were added. The memory was also showed in the management interface in a plain manner, without highlighting the potential private information, resulting in no participant proactively noticing the privacy inside memories. The past input were even not showed in a structured panel, leaving no participant aware of the potential privacy leakage. Although all participants re-vealed private information could be directly contained in the past input or memory during generation, 37/40 participants also envisioned private information could be inferred from multiple past inputs or memories. Worse still, 28/40 users were totally un-aware of the memory usage. Even these 28 participants who were aware of the memory usage, they could not understand or guess out how the memories are leveraged and integrated. Participants also has no manner to effectively regulate or modify the memory. The management panel is hard to find, and 35/40 participants never clicked it. The 5 participants who clicked it exhibited simple behaviors. Three participants directly deleted all the memories because of the memory risk, whereas two participants chose to retain all the memories regardless of the privacy risk. Participants typically commented, \textit{``I never considered GPT has this function before this day. I am definitely not known about the private information inferred from the memory.''}

\subsubsection{F2: Users Lack the Privacy Awareness towards the Memory Mechanism}


Despite the fact that users were un-aware of the long-term memory generation, they were also un-aware of the private information contained or potentially inferred from the past input and the memory unless explicitly told. 30/40 participants would never notice the private information in the past input before the experiment, and unfortunately, 36/40 participants never notice the private information in the memories before the experiment. They were even more un-aware of the risk that the private information could be inferred through combining different past user input and memory, which was exactly done in the usage of the long-term memory. In fact, among the interviews, all participants echoed \textit{``I have never thought about these information usage and leakage patterns before.''} All participants were never aware of the private information inferred combining the past inputs and memories. However, interestingly, participants could understand the inference process after explicitly told. This further un-veiled the in-transparency of the system's memory management.


\subsubsection{F3: Users Need the Control of the Memory Mechanism}

All participants reflected they lacked of control over the memory mechanism, highlighting the necessity for proactive memory management. All participants expressed the need for greater autonomy in controlling their memory, which can be categorized into three key actions: editing, adding, or deleting memories. Notably, a considerable percentage of participants identified privacy as the primary motivator for managing their memories, with privacy concerns being the most critical factor (35/40), followed by the accuracy of the stored information (14/40). 36/40 participants commented \textit{``I definitely need the system to provide me with the proactive modification permission.''} A few participants also hoped besides proactive modification, the system could automatically help them manage private information. When it comes to private information, users indicated the need to make decisions about deleting such data based on the usage, importance, and potential privacy risks associated with the memory. Specifically, categories such as health information, academic information, and personal basic information were frequently deleted due to their sensitivity, whereas categories like preference, formatting options and research interests were often retained to enhance usability.

\section{Design and Implementation of MemoAnalyzer}

MemoAnalyzer is designed to address privacy concerns aside users' tasks. It analyzes user-added memories and inputted information, providing a visualization for users to review, modify, or delete the data. This process is done before the real fusion of different memory and inputted information source during hypothetical later input (usually during training). Deletions are performed post hoc, striking a balance between preserving model output performance and safeguarding privacy, as the user's task has already been completed at that point.

\subsection{Design Goals}\label{sec:design_goal}

We proposed several design goals according to the formative study: 

\noindent \textbf{DG1: Enhance the transparency of (private) memory information inferred from user input and past memory through visualization. (F1)} Users were seldom aware of the private information inference, especially combining past input and memory together. Visualizing is an effective method to provide information to users intuitively~\cite{dasgupta2013measuring}. 

\noindent \textbf{DG2: Increase users' awareness of (private) memory inference through highlighting the inferred text and the original keywords. (F2)} Few users were aware of the memory inference process, especially the keywords involved in the inference. Additionally, the private information is not transparently illustrated. Thus, the visualization of the keywords could make users aware of the opaque process~\cite{9258413}. 

\noindent \textbf{DG3: Support users' proactive control of (private) memory information such as modifying and selectively adding. (F3)} Often, the system could automatically conduct the modification. However, towards the privacy content, users needed to have the proactive control, which users could selectively protect their privacy according to their preference and further propelled the forming of privacy awareness~\cite{ghaiumy2024personalizing}.

\subsection{Interface and Interaction Flow}\label{sec:interface}

\begin{figure}[!htbp]
    \centering
    \includegraphics[width=0.8\linewidth]{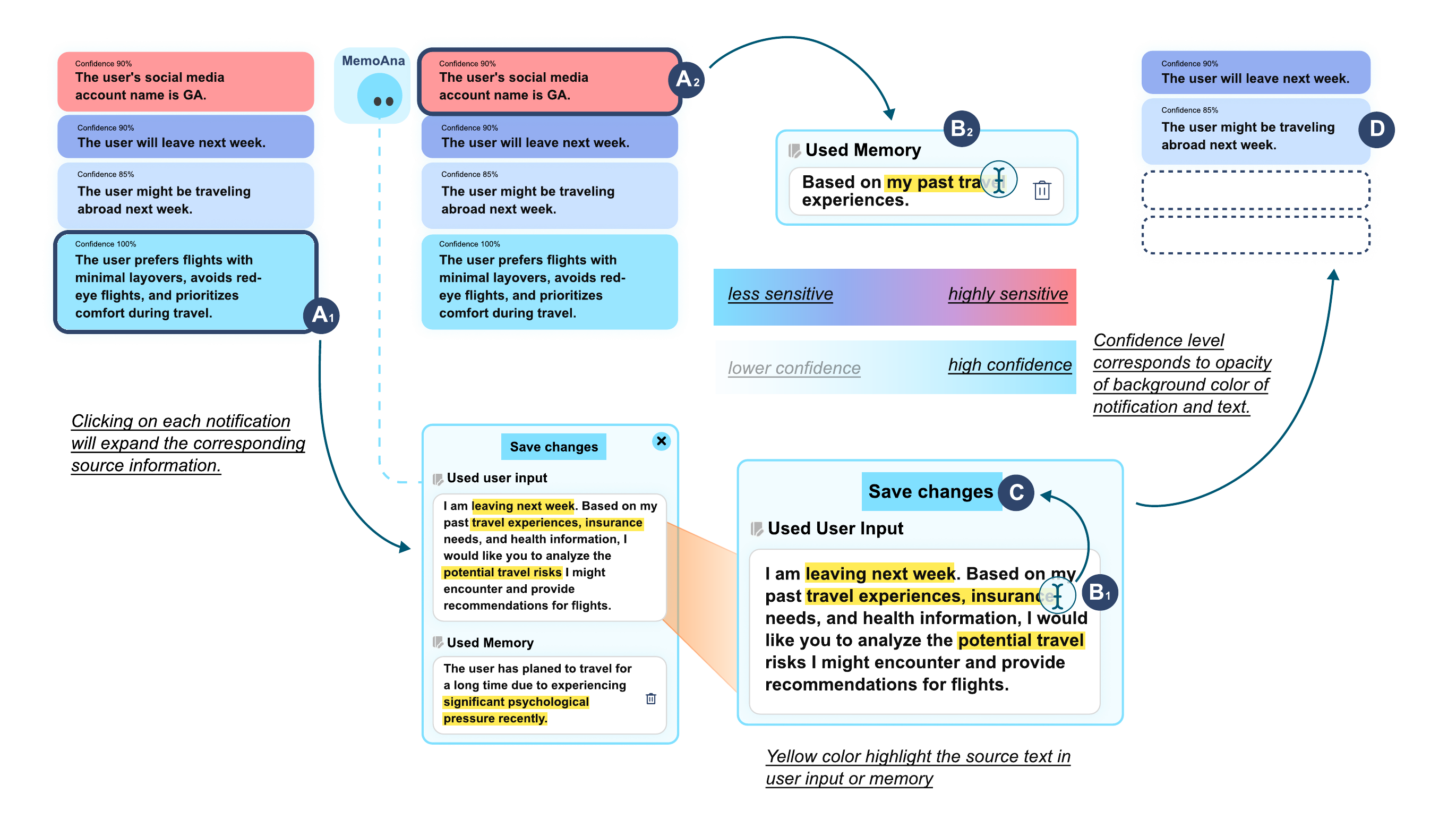}
    \caption{MemoAnalyzer's different functions. (A) The user click the notification attracting their curiosity. (B) The past inputs and memories used to infer the private information are expanded below. The specific phrases used for inference are highlighted to facilitate users' modification. (B1, B2) Users can edit or delete the memories while edit the past input. (C) User clicks the ``save changes'' button after modification to save the changes. (D) The inferred private information disappears or changes after user's modification.}
    \label{fig:interface}
\end{figure}

MemoAnalyzer adopted a minimal interface design~\cite{kim2009designing} in the form of a pop-up, as shown in Figure~\ref{fig:interface}. It followed the previous intervention design, placing the interface in the peripheral (the left) of the interface, reducing the disturbance to users' interaction. Upon completing the input, the users could view the private information inferred from the past inputs and memory histories, as demonstrated on the left of the interface (see Figure~\ref{fig:interface} (A)). The main elements in the intervention interface are the inferred private information based on the past input and the memory. We chose to demonstrate all inferred private information to increase users' control-ability and enhance their privacy awareness. MemoAnalyzer used different highlight according to the confidence of the inference and the sensitivity of the information. The confidence of the inference was defined as ``how sure the LLMs think the inferred information is?''. The confidence of the inference $c \in [0, 1]$ controls the transparency of the block, while the sensitivity of the information $s \in [0, 1]$ controls the color of the block. The more sensitive the information, the redder the information was highlighted. The relative privacy was determined through a questionnaire-based rating pilot study with the methods similar to the past literature about online text privacy~\cite{bhatia2018empirical,hanson2020taking}. We transformed the original mean ratings to between 0 and 1 using linear mapping. The color was determined by the following equation: \begin{equation}
    rgba(c, s) = (109 + s * (255 - 109), 172 + s * (117 - 172), 255 + s * (117 - 255), c) \end{equation}
We determined not to show the original text by default to reduce users' mental load of viewing the privacy information. Only when users proactively click the private information would the system show the information source to users. To reduce the cognitive load of users, we designed a hierarchical interaction approach. Notifications are given first, followed by detailed expansion. Clicking on the private information inferred would unfold the original text source and users could view the input and memory history (see Figure~\ref{fig:interface} (B)). The information source was divided into the past input (short-term memory) and the past memory (long-term memory). The past memory supported editing and deleting, while the past input supported editing. The keywords used to infer the private information were highlighted in yellow color (see Figure~\ref{fig:interface} (D)). We used the direct click to indicate the start of the editing, aligning with previous literature~\cite{zhu2017cept}. Users could edit on the history, delete or edit the memory and click ``Save Changes'' to record the modification (see Figure~\ref{fig:interface} (D)).

\subsection{Implementation of MemoAnalyzer}


We implemented MemoAnalyzer using Javascript and Python, where the main notification floating window is implemented as a plug-in. The frontend and backend adopted Flask framework. For the memory management and the inference, we used a one-shot prompting method, similar to GPT's memory generation and custom instruction processes~\cite{openai_custominstructions, openai_memory}. This approach requires no user data training, ensuring participants' privacy. The implementation is detailed in the following sections, \textit{with additional information provided in the supplementary materials}.

\subsubsection{Privacy Inference}

To better protect users' privacy, we used the most advanced method of inference~\cite{staab2024beyond}. We inferred the potential private information based on all the past user input in the current dialogue and all the past memories. The privacy inference required the system to ``infer and identify the personal sensitive information'' from ``the past inputs and the memory'' as much as possible. We also added prompts facilitating sensitivity highlighting and source tracking, which we detailed in the following sub-sections. We also added the prompts to reduce repetitive private information. We prompted the LLMs (using GPT-4o-2024-05-13) to output in a structured manner and extracted all the private information items to demonstrate on the interface. The input consisted of the step-by-step description, the definition of different private information types, rules, formats (including the private information, its type, the confidence, the original past inputs, the original memory) and the one-shot example. The one-shot example was manually crafted by experimenters. The structured output is a list of multiple private information, along with their confidence, types and the original text. 

\subsubsection{Sensitivity Highlighting}

We opted to output the confidence and sensitivity of the information alongside the inference. This choice is driven by 1) the latency would be reduced to a simple query, and 2) the LLMs were tested capable of handling the information and sensitivity together. The confidence was prompted to output with the private information. We also prompted the LLMs to output the sensitive information type. LLMs was asked to output the corresponding type along each private information. We used the sensitivity rating in the literature~\cite{bhatia2018empirical,hanson2020taking} to represent the sensitivity of the specific private information, and visualized them through the transition from red to blue.

\subsubsection{Source Tracking}

We used a prompt-based method alongside the previous steps for tracking the source of the information. We let LLMs to output the source information used to infer the privacy and especially the keywords used for inferring the privacy. The LLMs was asked to separately output the past input and the memory, along with the keywords. We also tagged the unique identifier of each input and let LLMs select the unique identifier to facilitate later modification and replacement. 

\subsubsection{Editing Proxy}

We proxied users' editing. Users edited on the left of the screen and the corresponding input as well as the memory would be modified correspondingly. This was achieved through assigning a unique id for each input, as mentioned in the last sub-section. Upon users submitting the modification, we tracked the modification and replaced the original history and the memory according to the unique identifiers through searching. This largely reduced users' time of viewing through all the past histories and modify manually.

\section{Study 2: Evaluating MemoAnalyzer}\label{sec:study_2}

We conducted a five-day in-lab study to evaluate the effectiveness of MemoAnalyzer compared with other memory management methods. 

\subsection{Participants and Apparatus}

We recruited 36 participants (12 males, 24 females) with a mean age of 23.9 (SD=2.7) from the XX campus (anonymized for submission) through snowball sampling \cite{goodman1961snowball}.  Participants reported moderate familiarity with LLM products (M=4.33, SD=0.47) and usage frequency (M=3.66, SD=1.24), but lower familiarity with AI techniques (M=3.33, SD=0.47), indicating regular use of LLMs but limited understanding of underlying AI technologies. Participants were also not familiar with privacy and security researches and techniques. No participant dropped the experiment and each participants received \$15 as compensation. The experiment was conducted through an online platform provided as a web service through Flask. Participants used their own laptop to connect to the provided website for the experiment to better mimick their own usage case. Experimenters connected to participants through online meeting\footnote{\url{https://meeting.tencent.com/}}. 


\subsection{Experiment Design}

Since the memory patterns of LLMs do not fully emerge in a single day, we conducted a five-day study, focusing on short-term memory each day and analyzing long-term memory across the entire period. The period aligns with the previous literature~\cite{zassman2024mindful,lee2024gazepointar} and is proved to be efficient by the results see Section~\ref{sec:study2_result}.

\begin{figure}[!htbp]
    \includegraphics[width=\textwidth]{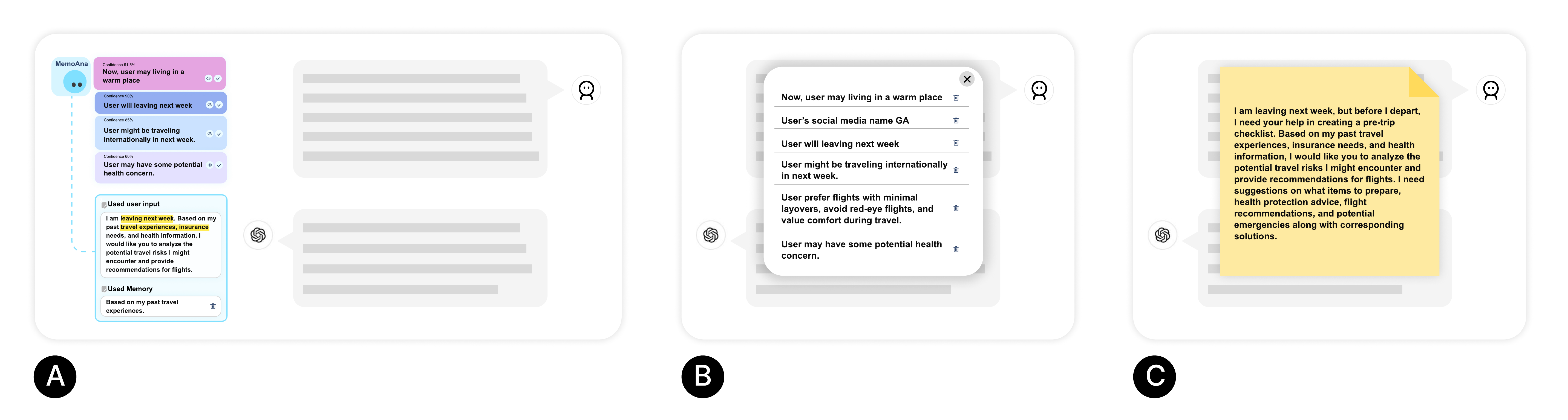}
    \caption{The experiment platform of different techniques, (a) MemoAnalyzer, (b) GPT-4o, (c) Manual. }
    \label{fig:experiment_platform}
\end{figure}

We used a one-factor within-subjects design with \textbf{technique} as the only factor. We compared MemoAnalyzer with two other techniques (see Figure~\ref{fig:experiment_platform}):

\begin{itemize}
    \item MemoAnalyzer: we implemented MemoAnalyzer according to the design and implementation section. Besides, we implemented the memory management interface as GPT products.
    \item GPT: we implemented the memory mechanism with a similar manner to GPT products\footnote{\url{https://chat.openai.com}} following official guidelines~\cite{OpenAI2024}. We used GPT to extract memory, to guarantee both extracting effect and a low latency. We implemented the same memory management interface as memoAnalyzer.
    \item Manual: we implemented the manual baseline with no context and memory, but a clipboard for users to manually copy and paste their past input. This method mimicking Temporary Chat~\cite{openai_Tamparychats}. They could also save the memory in the system similar to their local memorandum.
\end{itemize}

For all techniques, the back-end API was GPT-4o (version: GPT-4o-2024-05-13), the most advanced to facilitate comparison. We implemented the interface to support creating, managing dialogues for all systems. The chat history would be maintained, however the context would never be used for the Manual baseline. For the memory mechanism of all baseline techniques and MemoAnalyzer, we followed the guidance of OpenAI for implementing. For the parts that OpenAI did not specify, we detailed in supplementary materials for its implementation following the past guidance in the literature \cite{yen2024memolet}. We envisioned our memory mechanism is representative of the main stream LLMs and facilitate fair comparison. \textit{The detailed content and prompt of the implementation were shown in the supplementary material.} 

We selected three types of tasks related to users: work-related, life-related, and study-related~\cite{zhang2024s}. Each task type was designed and distributed based on the ShareGPT90K dataset\footnote{\url{https://huggingface.co/datasets/liyucheng/ShareGPT90K}}, commonly used for analyzing human-AI interaction. Tasks were carefully structured to be easily completed daily and to maintain correlations across days. Detailed task content is provided in supplementary materials. Participants were instructed to pseudo-anonymize their inputs to avoid sharing private information and mitigate potential ethical issues. Anonymization examples were provided for clarity.

\begin{figure}[!htbp]
    \includegraphics[width=0.8\textwidth]{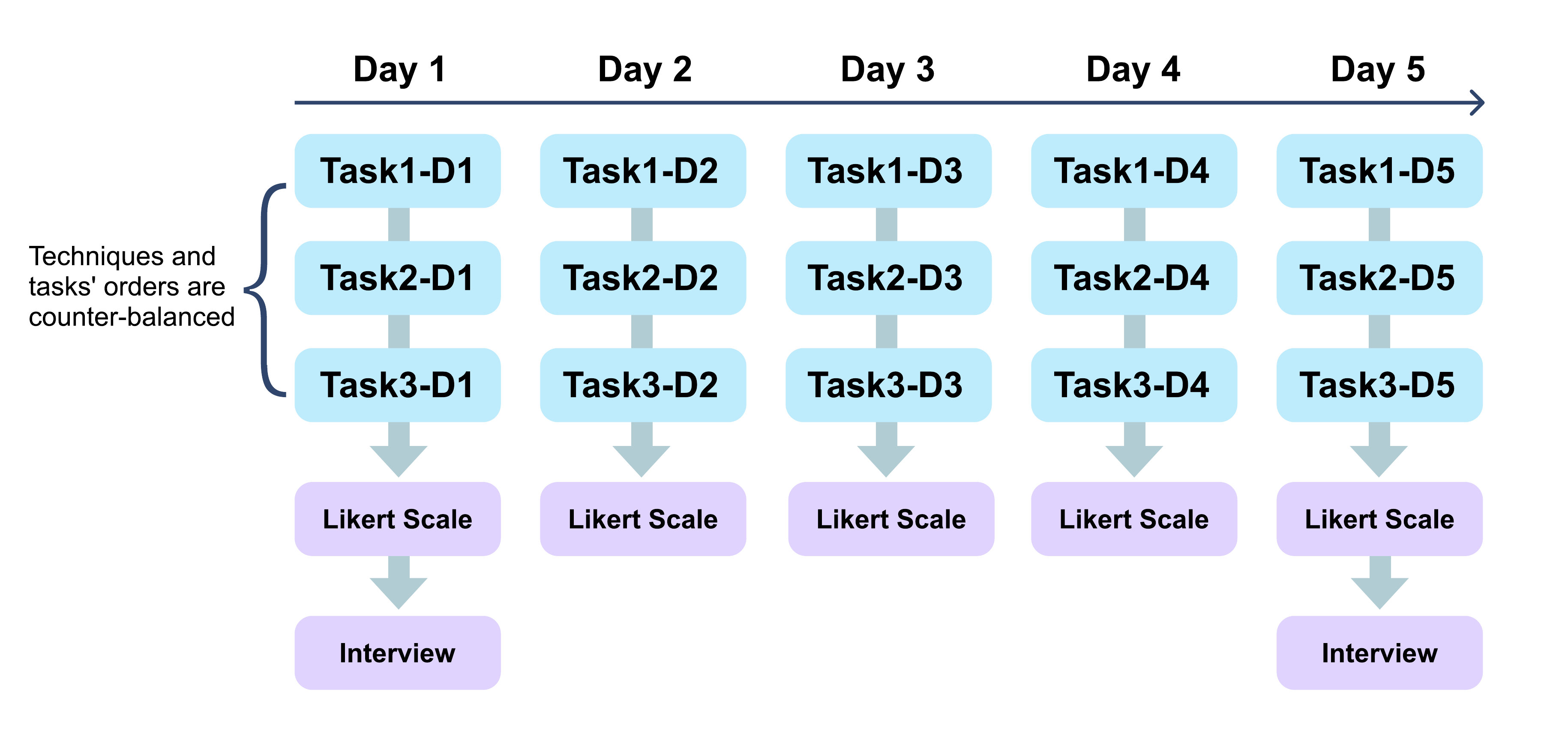}
    \caption{An overview of the study's process. There are interviews on Day-1 and Day-5 separately and questionnaires after each day's tasks.}
    \label{fig:study2_process}
\end{figure}

Figure~\ref{fig:study2_process} showed the experiment design. Each participant completed three five-day tasks, varied by technique and task type, all within a single scenario. Scenarios were counterbalanced among participants. Tasks were divided to reflect daily usage patterns, ensuring each task correlated with previous ones and remained manageable for daily completion.

We used questionnaires and also analyzed users' behaviors, with measurements shown in Table~\ref{tbl:metric}. We also conducted 15-minute semi-structured exit interviews to all participants after the experiment of Day-1 and Day-5 (see Figure~\ref{fig:study2_process}). The preset interview questions in the exit interviews included to let participants describe 1) how they use each functions in different systems to manage their memory and 2) how they perceive these functions regarding their usefulness and ease of use. The details of the exit interview was shown in the supplementary materials. Additionally, we asked participants about potential improvements of different techniques.

\begin{table}[ht]
    \centering
    \caption{Subjective and objective evaluation metrics for Study 2. The objective metrics were highlighted in blue. }
    \label{tbl:metric}
    \begin{tabular}{ll}
        \toprule
        \textbf{Category} & \textbf{Metrics} \\
        \midrule
        \colorbox{cyan!20}{\textbf{Efficiency}} & \begin{tabular}[t]{@{}l@{}}
                                \colorbox{cyan!20}{\textbf{Total completion time}}: time taken by users to complete tasks with each system. \\
                                \colorbox{cyan!20}{\textbf{Privacy management time}}: time taken by users to manage their privacy. \textit{There is no }\\
                                \textit{management time for Manual group.} \\
                                \textbf{Perceived privacy management speed}: whether the system could fast and  efficiently \\ collaborate with you in managing the privacy. \textit{There is no perceived privacy management}\\ \textit{ speed for Manual group.} \\
                              \end{tabular} \\
        \midrule
        
        \colorbox{cyan!20}{\textbf{Function usage}} & \begin{tabular}[t]{@{}l@{}}
                                     \colorbox{cyan!20}{\textbf{Frequency}}: measure how frequently participants used different functions \\
                                     such as selectively add, modify, delete the memories during memory \\
                                     generation and usage process.
                                   \end{tabular} \\
        \midrule
        \textbf{User satisfaction} & \begin{tabular}[t]{@{}l@{}}
                                       \textbf{Usability}: user satisfaction, ease of use and overall user experience. \\
                                       \textbf{Control}: whether users think they were in control of the private information. \\
                                       \textbf{Transparency}: whether the system transparent demonstrate the information to users. \\
                                       \textbf{Effectiveness}: whether the system could effectively handle privacy issues. \\
                                     \end{tabular} \\
        \midrule
        \textbf{Cognitive load} & \begin{tabular}[t]{@{}l@{}}
                                     \textbf{NASA-TLX}: NASA Task Load Index, cognitive load on users when \\
                                     interacting with the memory system.
                                   \end{tabular} \\
        \bottomrule
    \end{tabular}
\end{table}

\subsection{Procedure}

We first informed participants of the experiment and gave them 3 minutes to become familiarize with the experiment platform. They then needed to sign the user consent before proceeding the study. We detailed the potential harm in the user consent and they were informed they could quit the experiment at any time. They needed to complete 3 sessions of experiment differed by the technique managing their memory and private information. Within each session, they needed to interact with the system for several turns until they were satisfied with their answer. They input the question in the input area of Figure~\ref{fig:experiment_platform} and clicked the submit button to see the private information processed in the same area in MemoAnalyzer. They followed the experiment guideline for other systems. They needed to complete their corresponding part of task for each technique for each day across five days. After the completion of each technique the participants needed to fill in the subjective evaluation questionnaire. All techniques were video-recorded. After Day-1 and Day-5 we asked participants to participate in a semi-structured interview, in which participants also needed to comment on how and why they handle the private information. The experiment for each day lasted no more than 40 minutes and participants were each compensated 350RMB in total.

\subsection{Results}\label{sec:study2_result}

We conducted statistical testing to all behavioral and subjective rating data. We performed Repeated Measures Analysis of Variance (RM-ANOVA) and Tukey post-hoc comparisons to behavioral data, whereas we performed Friedman non-parametric tests and Nemenyi post-hoc comparison to subjective rating data. We further performed thematic analysis~\cite{braun2012thematic} to subjective interviews. The themes were generated through a combination of open-coding~\cite{khandkar2009open} and axial-coding~\cite{kendall1999axial}. 

\subsubsection{Privacy Protection Effectiveness}

\begin{figure}[!htbp]
    \centering
    \includegraphics[width=0.9\linewidth]{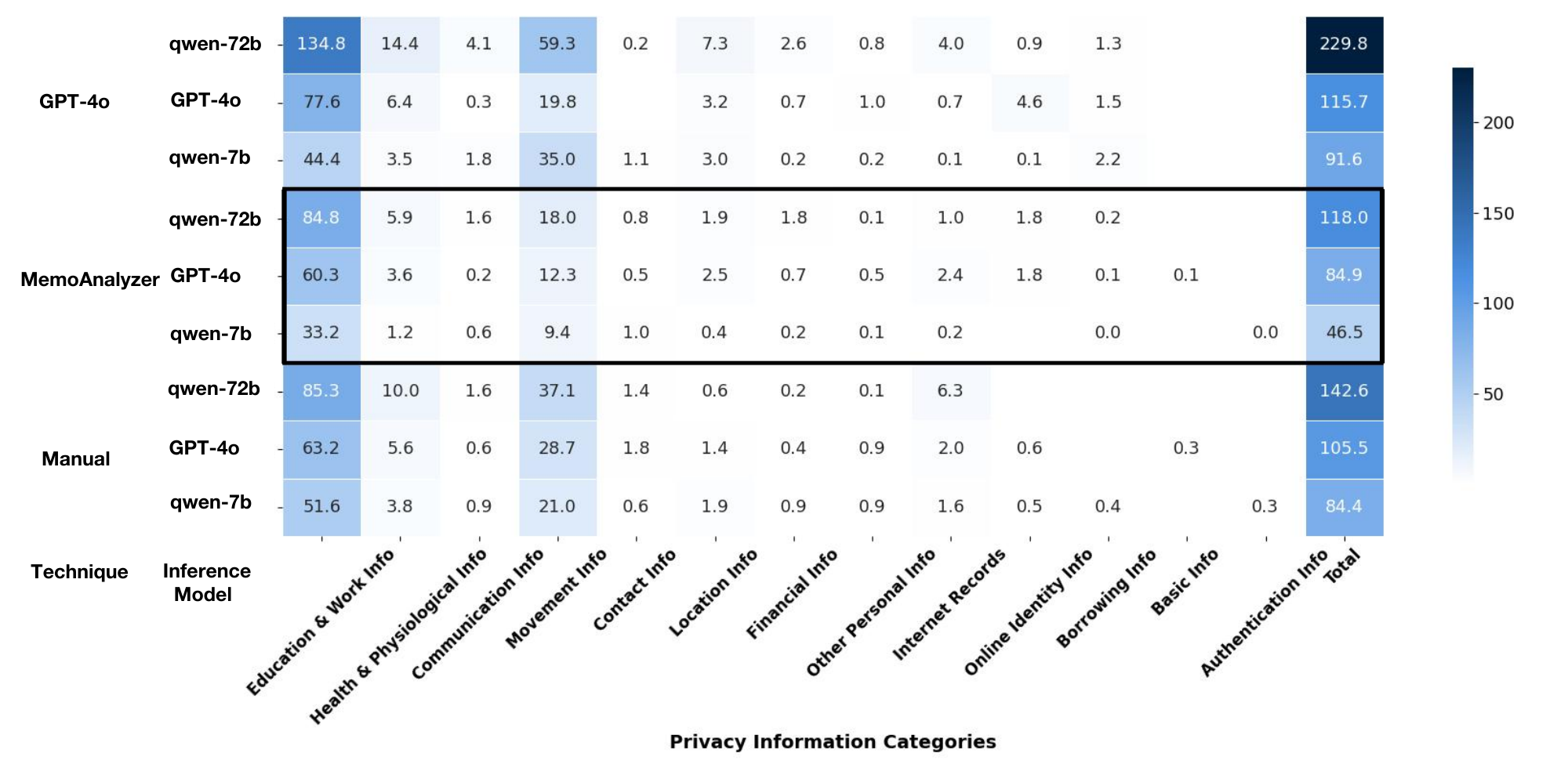}
    \caption{The heatmap of inferred information for different privacy information categories (the final column is the sum). The number denoted the private information item counts inferred using LLMs, averaged across participants. The horizontal axis denotes the technology (MemoAnalyzer, GPT-4o, Manual) and the LLMs used for inference (GPT-4o, qwen, qwen-7b). The numbers for MemoAnalyzer is outlined with black boundaries.}
    \label{fig:study2_privacy}
\end{figure}

In response to the threat model where users' past inputs and memories are used for training and inference in LLMs, leading to privacy leakage, we tested three commercial LLMs—GPT-4o (estimated $>$100B), Qwen-72B, and Qwen-7B—to infer private information from users' utterances and memory. On the fifth day, we collected participants' past inputs and memory histories, conducting five inference attempts to minimize random effects. Due to the model API limit, we randomly chose 25 participants from total 36 participants (accounting over the half) to conduct the analysis. Each dialogue was analyzed along with its related memory, and results were aggregated across dialogues. We reported the average number of inferred private data per participant, categorized and reported separately~\cite{bhatia2018empirical,hanson2020taking}\footnote{\url{https://www.tc260.org.cn/upload/2021-12-31/1640948142376022576.pdf}}. Figure~\ref{fig:study2_privacy} showed the number of different private information inferred using different LLMs. Specifically, for MemoAnalyzer, there was a marked reduction in the total amount of inferred private information. A statistical analysis revealed a significant difference in the total private inference for advanced models such as GPT-4o ($F_{2, 48} = 4.35$, $p < .01$, post-hoc $p < .05$, compared with GPT and $p < .05$ compared with Manual) and qwen-72b ($F_{2, 48} = 4.45$, $p < .01$, post-hoc $p < .05$ compared with GPT and $p < .05$ compared with Manual). Even compared with models such as qwen-7b, there is still a significant difference ($F_{2, 48} = 2.67$, $p < .05$), although post-hoc comparisons found no significant differences. This demonstrated the superior privacy protection capability of MemoAnalyzer. 

Participants were found to frequently input education and work information, indicated by its highest frequency. MemoAnalyzer has its pronounced effect in guarding participants' privacy, with over 22.3\% and 4.6\% percentage of private information reduction evaluating using GPT-4o. Besides, participants also input less private information in these less frequent categories, resulting in a less total number. 



\subsubsection{Interaction Time}
\begin{figure}[!htbp]
    \subfloat[Total time.]{
        \includegraphics[width=0.5\textwidth]{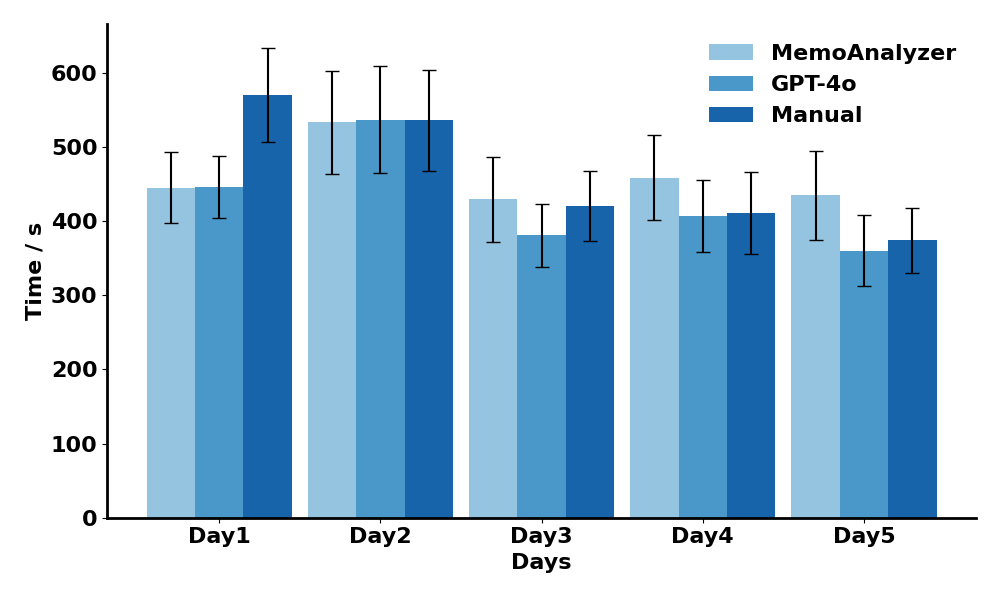}
    }
    \subfloat[Privacy protection time.]{
        \includegraphics[width=0.5\textwidth]{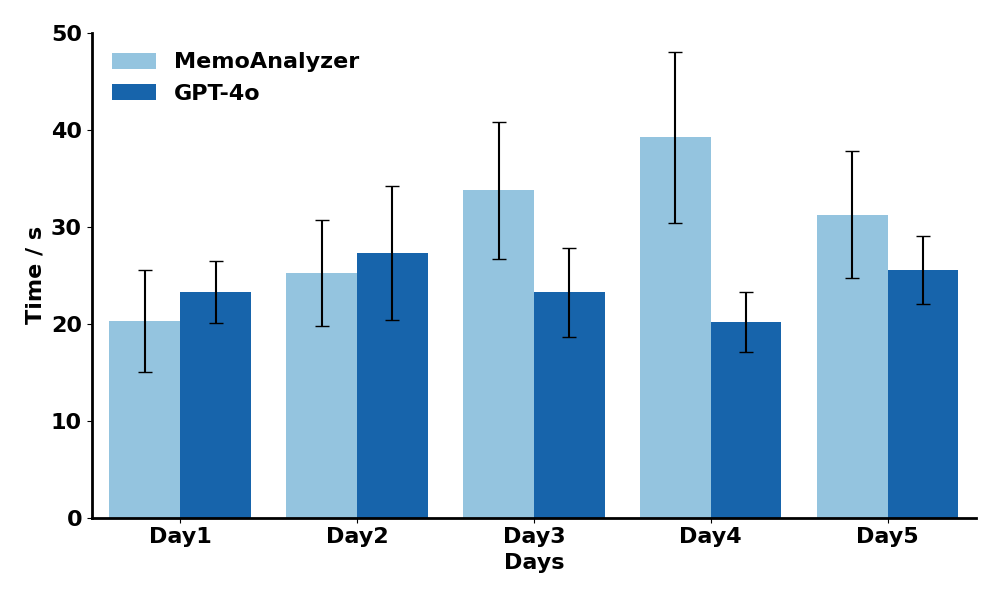}
    }
    \caption{(a) Total time and (b) Privacy protection time for each day for each technique. Errorbar indicated one standard error.}
    \label{fig:study2_time}
\end{figure}

We defined total time as the duration from the first character input to the moment the website was closed, indicating task completion. Privacy protection time was measured by participants’ interactions with notifications or memory panels for privacy management. Figure~\ref{fig:study2_time} presents the total and privacy protection times for each technique. Total time comprised both privacy protection and pure task completion time. \textit{MemoAnalyzer} exhibited comparable total times (M=460.3s, SD=58.6s) to \textit{GPT-4o} (M=426.2s, SD=50.6s) and \textit{Manual} (M=462.7s, SD=55.6s). Similarly, the privacy protection time for \textit{MemoAnalyzer} (M=29.9s, SD=6.6s) was close to \textit{GPT-4o} (M=23.9s, SD=4.3s). No significant differences in total time were observed across techniques, except on Day 1, where both \textit{MemoAnalyzer} and \textit{GPT-4o} outperformed \textit{Manual}.

Over the five days, time had no significant effect on the total completion time for \textit{MemoAnalyzer} ($F_{4, 140} = 1.32$, $p = .26$). In contrast, time significantly affected the total completion time for both \textit{GPT-4o} ($F_{4, 140} = 4.33$, $p < .01$) and \textit{Manual} ($F_{4, 140} = 5.01$, $p < .001$). This suggests that MemoAnalyzer maintained consistent task completion and privacy management times.


\subsubsection{Interaction Statistics}\label{study2:interaction}

We calculated the interaction statistics for each day for each technique, as shown in Table~\ref{tab:metrics}. 

\textit{Notify}: Notify refers to the average number of notifications per dialogue. MemoAnalyzer consistently achieved a high notification rate, indicating effective detection of private information requiring user control. The stable notification rate on Days 4 and 5 suggests that MemoAnalyzer effectively managed privacy, with participants disclosing less uncontrolled private information. 

\textit{Click}: Click refers to the average number of clicks per task. The click count initially increased, then decreased over five days, reflecting users' growing need for privacy control as concerns heightened, followed by stabilization as privacy risks diminished. This trend demonstrates MemoAnalyzer’s effectiveness, with privacy management stabilizing by Day 5. 

\textit{Revise}: Revise refers to the average number of revisions per task in MemoAnalyzer. Revision frequency increased over time, stabilizing on the last two days, indicating users' growing need to modify private information, with concerns reaching equilibrium as privacy control improved.

\textit{Revise(4o)}: Revise is defined as the average number of revise times per task for GPT-4o. For GPT-4o, the revision rate was consistently lower than MemoAnalyzer’s. Users found it challenging to identify and manage GPT-4o’s private information, resulting in fewer proactive deletions and a higher privacy risk.


\textit{Use Input}: Use Input refers to the average use of past input in each inference. The frequency of past input usage increased as participants incorporated more historical data into their dialogues, leading to greater reliance on past inputs over time, especially as dialogue histories accumulated.

\textit{Use Memory}: Use Memory refers to the average utilization of memory for each inference. The usage rate remained stable across the five days, which indicates the robustness and effectiveness of MemoAnalyzer.

\textit{Coverage}: As MemoAnalyzer would highlight the original text when inference, we defined coverage as the overlap of users' memory and the highlighted original text. Coverage remained high throughout the experiment, with a slight dip on Day 2 as participants experimented with modifications beyond the highlighted text, which proved less effective, returning the coverage to 96\% by Day 3.

\begin{figure}[!htbp]
\subfloat[]{
    \includegraphics[width=0.33\textwidth]{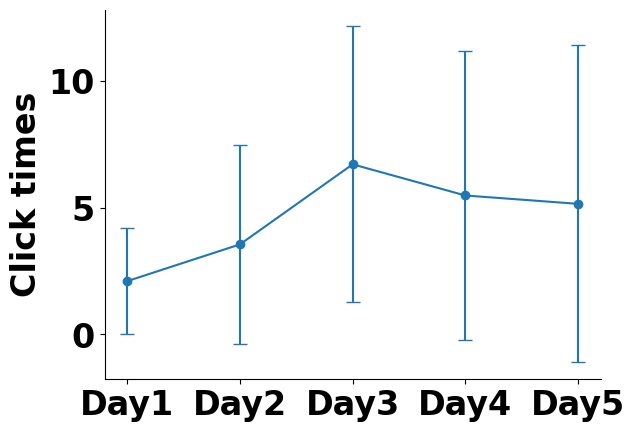}
}
\subfloat[]{
    \includegraphics[width=0.33\textwidth]{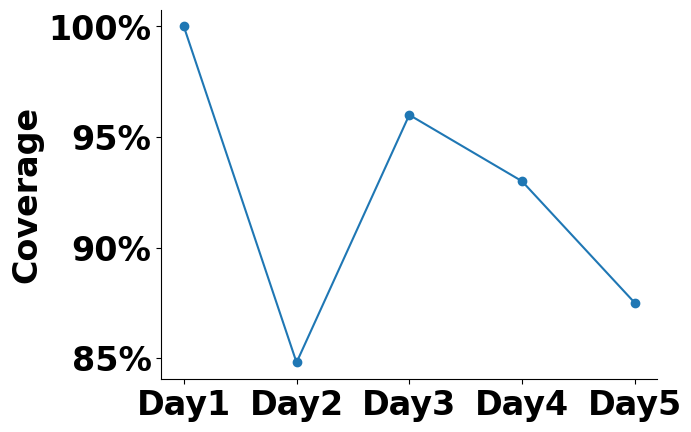}
}
\subfloat[]{
    \includegraphics[width=0.33\textwidth]{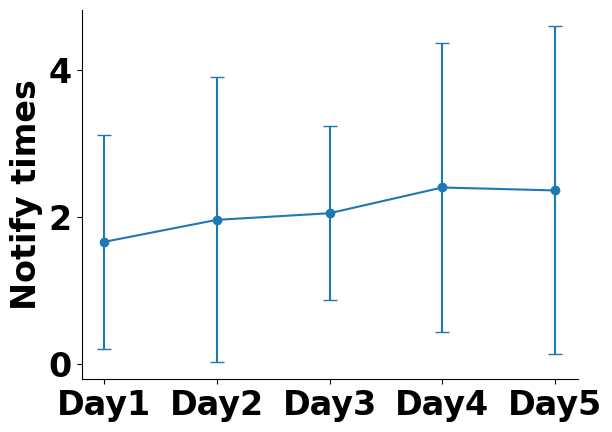}
}

\subfloat[]{
    \includegraphics[width=0.33\textwidth]{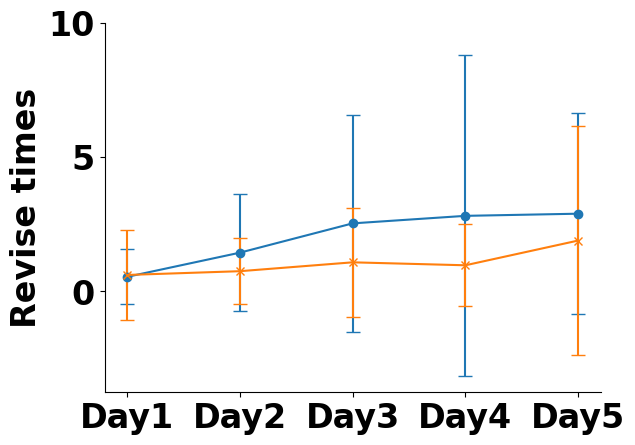}
}
\subfloat[]{
    \includegraphics[width=0.33\textwidth]{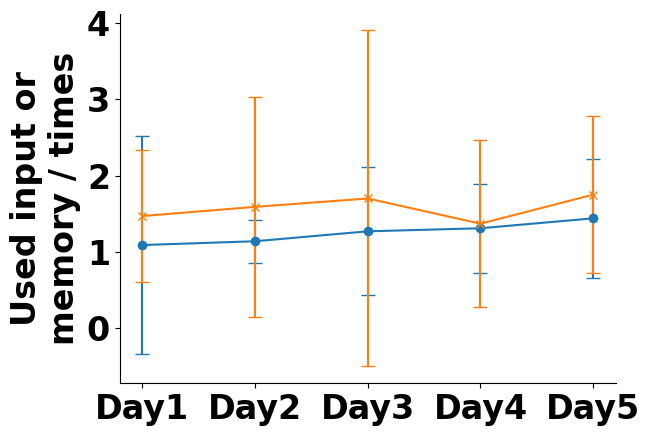}
}
\caption{Trends of different metrics over five days: (a) \textbf{Click}: number of user clicks on inferred information, (b) \textbf{Coverage}: overlap between users' memory and the highlighted original text, (c) \textbf{Notify}: number of notification pop-ups, (d) \textbf{Revise}: the orange line shows the number of user modifications or deletions of past inputs/memories for MemoAnalyzer, while the blue line shows deletions of past memories for GPT-4o, (e) \textbf{Use of input and memory}: the orange line shows the average number of memories used per inference, and the blue line shows the average number of past inputs used per inference. Errorbar in (a), (c), (d), (e) indicated one standard deviation. }
\end{figure}

The results demonstrate the effectiveness of MemoAnalyzer across various dimensions. Participants showed consistent modification patterns, typically adjusting their private information after viewing the answers, or during the model's output phase in the next round. This reduced overall task time. Two behaviors were observed: 1) clicking on high-risk privacy items, and 2) reviewing all inferred private information to examine the inference process. Participants primarily modified or deleted highlighted content, confirming MemoAnalyzer's ability to provide precise privacy control.

\subsubsection{Subjective Ratings}\label{sec:study2_subjective}

\begin{figure}[!htbp]   
    \subfloat[NASA-TLX.]{
        \includegraphics[width=0.85\textwidth]{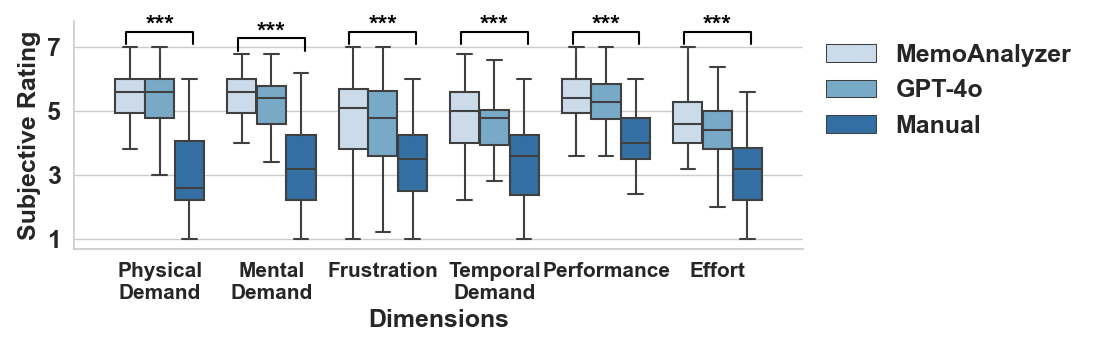}
    }
    
    \subfloat[Other subjective ratings.]{
        \includegraphics[width=0.85\textwidth]{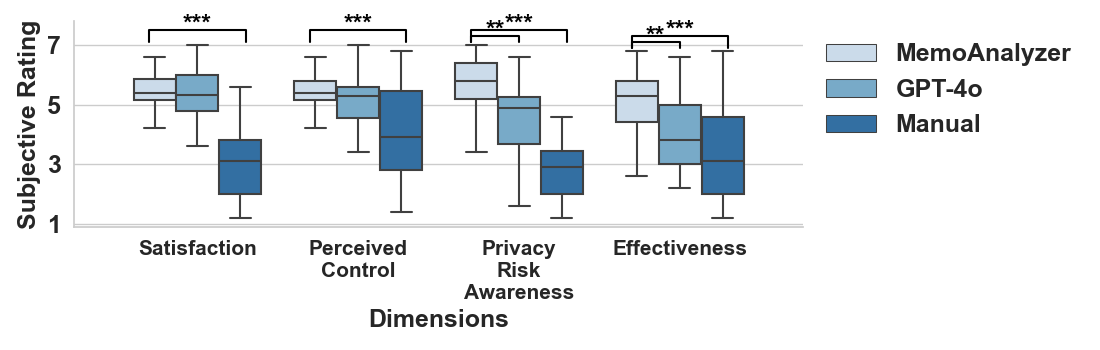}
    }

    \caption{The (a) NASA-TLX, and (b) other subjective ratings of all participants  (7: most positive, 1: most negative). Errorbar indicated one standard deviation. The significant differences between conditions were from post-hoc analysis.}   
    \label{fig:study2_subj}
\end{figure}

Figure~\ref{fig:study2_subj} showed the participants' subjective ratings. We found significant effects of techniques on all dimensions ($p < .05$). Notably, \textit{MemoAnalyzer} was praised as for its higher satisfaction ($\chi^2_2=51.6$, $p < .001$, post-hoc $p < .001$) and perceived control ($\chi^2_2 = 16.3$, $p < .001$, post-hoc $p < .001$) compared to the manual baseline. This demonstrated the effectiveness of MemoAnalyzer's collaborative design to provide users with control as well as maintaining the machine agency and accuracy in the same time. 

\textit{MemoAnalyzer} were further favored for its superior privacy risk protection effect ($\chi^2_2=46.9$, $p < .001$, post-hoc $p < .001$ compared with \textit{Manual}, $p < .01$ compared with \textit{GPT-4o}) and the effectiveness ($\chi^2_2=46.9$, $p < .001$, post-hoc $p < .001$ compared with \textit{Manual}, $p < .01$ compared with \textit{GPT-4o}) compared with \textit{Manual} and \textit{GPT-4o}. This demonstrates the system's strong ability to protect sensitive data while ensuring usability and speed. It highlights \textit{MemoAnalyzer}'s effectiveness in mitigating privacy risks, surpassing traditional methods, and aligning with our design goals by fostering privacy awareness through intuitive, user-driven control.

In terms of cognitive load and effort, \textit{MemoAnalyzer} reduced physical ($\chi^2_2 = 48.9$, $p < .001$, post-hoc $p < .001$) and mental demands ($\chi^2_2 = 39.0$, $p < .001$, post-hoc $p < .001$), frustration ($\chi^2_2 = 16.6$, $p < .001$, post-hoc $p < .001$), temporal demand ($\chi^2_2 = 18.6$, $p < .001$, post-hoc $p < .001$), performance ($\chi^2_2 = 29.8$, $p < .001$, post-hoc $p < .001$) and effort ($\chi^2_2 = 29.9$, $p < .001$, post-hoc $p < .001$) compared with \textit{Manual}. These proved \textit{MemoAnalyzer} was easier to use and could manage users' privacy easily. 


\subsubsection{Subjective Comments}
\textbf{Timely Reminding and reflection:}
29/36 participants commented the notification could help them gain privacy awareness better through visualization of the private information. In particular, they gave high comments for the confidence level and the information sensitivity visualization design. P15 commented that \textit{''I used to first look at the content of the notification, then pay attention to the color of the pop-up box. Brighter colors catch my attention more easily. The confidence percentage and color transparency help me realize how accurate the information is.''} 
28/36 participants found the notification help user reflect, some privacy information user did not notice originally will lead to privacy leaks. In all notifications, 26/36 participants give priority to relevant with themselves and correct personal information.

\textbf{Increasing Privacy Awareness:} We also found the increase of participants' trust towards the MemoAnalyzer(19/36 participants), which echoed the previous literature that memory management~\cite{yen2024memolet} and transparent design~\cite{reinhardt2021visual} could increase participants' trust. They thought the information origin and the visualization of the important information also help them modify the private information faster. \textit{``I find this plugin very helpful, as it enables me to identify where privacy may be exposed in my conversations. This allows me to improve my dialogues while managing my privacy more effectively.'' (P23)} P12 also thought that \textit{``It effectively highlights information I unintentionally reveal that could be inferred by the model.''} After 5 days of the experiment, 24/36 participants understood the meaning of private information better than before the experiment.

Participants also gave constructive comments. For example, P25 commented that \textit{``When using these two technologies, I find it difficult to be aware of the need to manage private information most of the time. When reviewing memory in Baseline-GPT, I adopt the same privacy management strategy as in MemoAnalyzer. While using Baseline-Manual, I try to add only content that does not contain private information; however, if certain private details must be entered, I will delete them after completing all tasks.''}. P10 also argued that \textit{``The advantage of the plugin is that it clearly identifies information I unintentionally reveal, which could be inferred by the model. However, its downside is the low frequency of use in my tasks, which may require extended usage to accumulate more text for inference.''}

\section{Discussion and Design Implications}

\subsection{Emerging Issues of Private Memory Information}

The emerging privacy risks associated with language models, such as membership inference attacks~\cite{staab2024beyond} and data leakage attacks~\cite{pan2020privacy}, have heightened concerns in the privacy landscape of human-AI interaction. The persistence of memory in LLMs further exacerbates these challenges by intensifying the risk of data leakage. Participants were generally unaware of the privacy risks associated with RAG memory and were unfamiliar with its implications. While they recognized the existence of context memory, they lacked understanding of how to mitigate associated risks, complicating the integration of memory into LLMs. Despite promises from companies like OpenAI to exclude private information unless necessary, participants often disclosed sensitive data that was required for subsequent tasks~\cite{zhang2024s}, highlighting the need for effective management. Our work represents the first attempt to address memory-related privacy concerns, leaving place for future exploration in quantitative regulation and algorithmic governance of memory systems~\cite{yen2024memolet}.

\subsection{Designing and Managing Private Information in Memories}

Memories are essential in human-AI interactions, enhancing personalization by retaining past exchanges. While the risks of using memory data for training are similar to those of using direct user inputs, the threats are more severe because memory data is retained longer, increasing the potential for misuse. We further found users seldom noticed and cared the existence of memory (see Section~\ref{sec:study1_result}). This paper presents the first framework for managing private information in AI memories, specifically addressing vulnerabilities to membership inference attacks~\cite{pan2020privacy} and privacy leakage~\cite{staab2024beyond}. Previous research~\cite{yen2024memolet} has addressed memory usage in AI systems but has overlooked the private information within memories and failed to model users' mental constructs regarding privacy. Recognizing that not all memory information is essential, we propose a collaborative method involving both users and large language models to select, retain, and utilize memory data according to users' preference~\cite{asthana2024know}. This approach aims to optimize the balance between leveraging valuable memory information and safeguarding user privacy.

\subsection{Feasibility of MemoAnalyzer} 


In this paper we validated the MemoAnalyzer in daily usage cases, where MemoAnalyzer reached better privacy protection effect with comparable time cost. MemoAnalyzer is also expected to be far better given the extreme usage case where private information are ample. We found in this extreme case participants could complete time with a total time comparable with the GPT and the manual baseline. Additionally, participants could control the privacy to 77.7\% of the GPT baseline and 95.4\% of the manual baseline calculated by number when inferred with GPT-4o. We envisioned MemoAnalyzer as the first system in effectively controlling the memory both no matter in the realistic setting or ``cold start'' setting (as on Day-1) or in an extreme setting (in Day-3 to Day-4), which could be adapted to various LLMs tasks such as co-writing~\cite{yuan2022wordcraft} and ideation~\cite{liu2024ai}.

MemoAnalyzer adopted the reactive setting~\cite{deng2024towards} for human-AI collaboration. This enabled participants the proactive control which increased their agency \cite{hwang2022ai} and enhanced their privacy awareness (see Figure~\ref{fig:study2_subj}). Participants were found to proactively determine which information to click and select the private information to control according to their preference (see Section~\ref{study2:interaction}). The reactive design also enabled participants to check before the system automatically delete their important information, which balanced performance enhancement and privacy protection (see Figure~\ref{fig:study2_privacy} and~\ref{fig:study2_time}). 

\subsection{Design Implications}

We proposed several design implications for the memory management of LLMs to fertilize future design.




\textbf{Balancing Control and Efficiency in Private Memory Management Through Proactiveness Levels.} We used hierarchical design to balance the control and efficiency, where users could click and collapse to view the original source of private information inference result. This allowed the completion of main task while maintaining efficiency for privacy protection. Hierarchical design in privacy protection could also be leveraged in privacy and security settings on smartphones~\cite{bourdoucen2024privacy} and text anonymization~\cite{staufer2024silencing}, where multiple levels of information or information source needed to be demonstrated efficiently and in the same time reduce users' mental load. 

\textbf{Visualizing the Source of the Privacy Risk to Foster Privacy Awareness.}
MemoAnalyzer enhance user privacy awareness by visually displaying how inferences are made from stored memories, which similar systems in online chatting~\cite{staab2024beyond} and personalization~\cite{asthana2024know} scenarios could also adopt. A clear, intuitive visualization of the data relationships and inferences, perhaps through highlighting key phrases or linking them to past user inputs, will help users understand how their personal information is being processed. This transparency can promote informed decision-making regarding which memories or information to retain, modify, or delete, ultimately fostering greater trust in the system’s privacy protections.

\textbf{Assigning Distinct Roles to Users and AI to Enhance Agency and Efficiency.}
Assigning users and AI different roles in the privacy management process can significantly improve both perceived agency and efficiency. For instance, users could focus on indicating their preferences or highlighting sensitive information~\cite{asthana2024know}, while the AI executes memory adjustments such as deletion, modification, or preservation based on those preferences. This role division reduces the cognitive burden on users, allowing them to concentrate on higher-level privacy decisions while ensuring the system efficiently handles the operational aspects of memory management.

\section{Ethical Considerations}
We acknowledged that our research may have ethical issues. We followed Menlo report \cite{bailey2012menlo} and Belmont report \cite{beauchamp2008belmont} in designing the studies and tried our best to avoid ethical concerns. In all studies, we compensated participants according to the local wage standard and told the participants at the beginning of the experiment about the potential benefits and harms. Participants were allowed to quit at any time in the experiment if they felt uncomfortable or for other reasons. Our experiment aimed at solving the privacy policy reading problems through designing applications to facilitate reading. The participants may potentially benefit from reading the privacy policy as they could acquire more information regarding their personal information collection. Besides, all the participants' experimental data was stored on a local device with encryption.

\section{Limitation and Future Works}
We acknowledged that our study have limitations and regard these as future directions. First, although we tried to diversify the background of the participants, we recruited the participants through snowball sampling in the campus, which restricted the age and educational background. University students was a group with relatively higher educational background than the average~\cite{betts1999determinants} and may understand the memory mechanism easier. This would further highlight the problems that memory mechanism is opaque and hard to understand.  Second, the experiment may face social desirability \cite{chung2003exploring} and recall bias \cite{coughlin1990recall} of the participants, during which participants may utter more opinions towards memory mechanism. We selected the most common problems of participants and envisioned these were real problems that needed handling.


\section{Conclusions}

This paper presents MemoAnalyzer, a proactive memory management system designed to mitigate privacy risks in human-LLM interactions. Our research highlights the opaque nature of memory mechanisms in current LLMs, which users are largely unaware of. Through a semi-structured interview and a five-day user study, we identified a significant gap in user awareness and control over long-term memory retention. MemoAnalyzer effectively addresses these issues by providing transparency, visualization, and user-driven control over private information, which was validated through improvements in privacy awareness, perceived control, and user satisfaction. This work contributes to the broader discourse on privacy-conscious AI design by demonstrating how user-centric privacy tools can enhance trust and control without impacting system performance. Future work can explore the scalability of MemoAnalyzer in more diverse real-world settings and investigate further improvements in user interaction with privacy management systems.

\bibliographystyle{ACM-Reference-Format}
\bibliography{sample-base}

\appendix 
\section{The Questions in Study 1}\label{app:question_study1}

The followings are the questions in Study 1. 
\subsection{Demographics}

1. What is your age range?

2. What is your highest level of education?

3. What LLMs have you used?

4. How well do you think you understand LLMs products? (self-evaluation, 1: very little, 5: very well)

5. How often do you use LLMs products? (self-evaluation, 1: less than once a year, 2: several times a year, 3: several times a month, 4: serveal times a week, 5: several times a day)

6. How well do you understand AI technology? (self-evaluation, 1: very little, 5: very well)

7. what is your field of study or education focus? (IT-related, privacy and security-related, design-related, others, multiple selections allowed)

\subsection{Memroy-related Questions}

1. How do you think the memory mechanism of LLMs works?

2. Are you aware that LLMs sometimes remember things you've said before? Can you provide a specific example?

3. What benefits do you think the memory function of LLMs can provide?

4. (Follow-up to the previous question) Have you experienced such benefits? How exactly, and why?

5. What privacy threats do you think the memory function of LLMs might pose? (Please elaborate)

6. (Follow-up to the previous question) Have you experienced such threats? How exactly, and why?

7. Are you willing to use the memory function of LLMs? Why? (You can discuss work or study examples with clear formats)

\textit{(Below is instruction) This memory mechanism discussed here works through external knowledge storage and retrieval, essentially following the Retrieval Augmented Generation (RAG) approach. During user input, the LLMs detects whether to store certain information and adds corresponding natural language or vectorized representations to the memory, which enhances the model's capabilities when used. The memory function can offer users a more personalized experience, with the information coming from user input.}

8. After this introduction, what privacy threats do you think the memory function or LLMs? 

9. How severe do you think these threats are? (1: very low, 5: very high)?

10. How do you think the memory function of LLMs should serve you? 

\subsection{Inference-related Questions}

1. What personal information do you think the following text reveals about you? 

2. How high do you think the privacy risk of this information is? 

3. If a LLM can infer this kind of personal information, what do you think the following text reveals about you?

4. How high do you think the privacy risk of this information is?

\subsection{Thoughts on Memory}

1. what are your expectations for managing AI memory? What are you core needs?

2. What features should an ideal AI memory management system have? How individual people control it?

3. Do you think it's important for an AI memory management system to clearly show the source of the memory (reasoning process, source dialogue)? Why?

4. Do you think it's important for an AI memory management system to clearly show how the memory should be used? How should it be presented during use?

\subsection{Thoughts on Inferences}

1. What are you expectations or thoughts about AI inferring personal information? What are you core needs?

2. Do you need the system to transparently show the inference process related to privacy? 

3. Would you need assistance in understanding what kind of information might be inferred?

4. What kind of visualization would you prefer for this?

5. What solutions do you think could address the issue of AI inferring personal information?

\end{document}